\documentclass[twocolumn,amsmath,amssymb,prl,aps,showkeys]{revtex4}
\usepackage{graphicx}
\usepackage{amsmath,amssymb}
\usepackage{mathrsfs}
\usepackage{color}
\usepackage{afterpage}
\usepackage[version=3]{mhchem}
\usepackage[normal]{subfigure}
\usepackage{natbib}
\usepackage{soul}

\begin{document}
\title{Elucidating the role of Sn-substitution and Pb-$\Box$ in regulating stability and carrier concentration in CH$_3$NH$_3$Pb$_{1-X-Y}$Sn$_X$$\Box_Y$I$_3$}
\author{Debalaya Sarker\footnote{debalaya.sarker@physics.iitd.ac.in} and Saswata Bhattacharya\footnote{saswata@physics.iitd.ac.in}}
\affiliation{Department of Physics, Indian Institute of Technology Delhi, Hauz Khas 110016, New Delhi, India}
\date{\today}
\begin{abstract}
We address the role of Sn-substitution and Pb-vacancy (Pb-$\Box$) in regulating stability and carrier concentration of  CH$_3$NH$_3$Pb$_{1-X-Y}$Sn$_X$$\Box_Y$I$_3$ perovskite using density functional theory, where the performance of the exchange-correlation functional is carefully analyzed, and validated w.r.t. available experimental results. We find 
the most stable configuration does not prefer any Pb at 50\% concentration of Sn. However,  
the Pb-$\Box$s become unfavourable above 250K due to the reduced linearity of Sn-I bonds. For n-type host the Sn substitution is more preferable than Pb-$\Box$ formation, while for p-type host the trend is exactly opposite. The charge states of both Sn and Pb-$\Box$ are found to be dependent on the Sn concentration, which in turn alters the perovskite from n-type to p-type with increasing $X$ ($>$0.5). 
\end{abstract}
\maketitle
\begin{figure}[t!]
\includegraphics[width=0.9\columnwidth,clip]{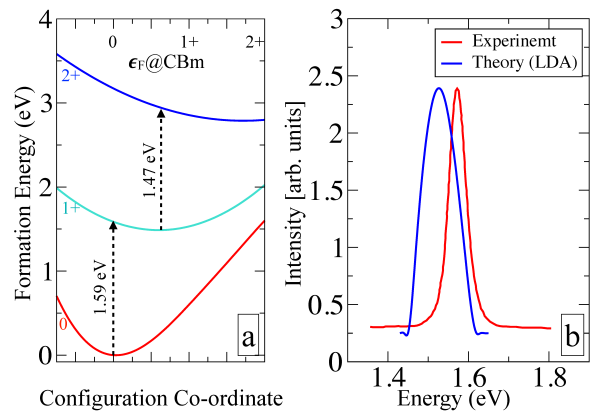}
\caption{(a) Configuration coordinate diagram for pristine MAPbI$_3$: The LDA formation energies in the three different charge states (0, +1, +2) are plotted as a function of the displacement of atoms. The smooth lines are obtained by fitting parabolas to the three data points of each curve. Since pristine MAPbI$_3$ is n-type in nature, fermi level ($\epsilon_\textrm{F}$) is aligned with the conduction-band minimum (CBm). (b) Calculated emission spectra from both theory and experiment. The experimental data is taken from Ref~\cite{Nat_PL_Michele}  and the employed theoretical methodology is explained in details in Ref~\cite{PRL_Rinke_MgO}.}
\label{func}
\end{figure}

%%%%%%%%%%%%%%%%%
Inorganic-Organic perovskites, mainly CH$_3$NH$_3$PbI$_3$ (MAPbI$_3$) and it's derivatives, are one of the very few materials that have created a global research sensation in just 2-3 years of their discovery~\cite{Kojima-JACS, Lee-Science, NatPhot-Green, NatMat-Gratzel}. 
Long diffusion length, high carrier mobility, suitable optical band gaps ($\sim$1.54 eV), strong absorption of light alongside of very cheap manufacturing costs 
have made this material a leading member of present solar and photovoltaic community~\cite{NatLet-Jeon-HighPerform, SciRep-Kim, Lee-Science, Yin-APL-defects, Science_diff-len-2015, InorgChem_MAPbSnI3_Constantinos}.
However, for an ideal light harvester absorbance of some ultraviolet to near-infrared photons (upto 1.1 eV) along with all visible lights of the solar spectrum is desirable~\cite{JACS_Feng_Sn-Pbmixed}. Reportedly, inclusion of Sn in the perovskite network (i.e. MAPb$_{1-X}$Sn$_X$I$_3$) can reduce MAPbI$_3$'s optical band gap to achieve this. Moreover, the presence of hazardous Pb has rendered MAPbI$_3$ from its practical applications in non-toxic perovskite solar cells.  Therefore, reducing the extent of Pb by substituting a suitable alternative metal (e.g. Sn, Ge, Sr, etc.) in the perovskite has also become crucial. However, the complete removal of Pb from the perovskite cage hampers its solar cell performance and stability hugely~\cite{JACS_Feng_Sn-Pbmixed, Sn_chg_Yukari_11, nature_rev_samuel_15}. In view of these major issues, over the years several attempts have been made to substitute Pb with Sn 
by forming a hybrid perovskite MAPb$_{1-X}$Sn$_X$I$_3$ ~\cite{JACS_Feng_Sn-Pbmixed, InorgChem_MAPbSnI3_Constantinos, SciRep_MAPbSnI3_Liu, RSC_MAPbSnI3-Kanhere}.%for its effective %%%%%%%%%%
\\ \indent Experiments have shown that the lowest band gap is achieved for $X$ $\geq$ 0.5-0.75~\cite{JACS_Feng_Sn-Pbmixed}. First-principles based calculations have predicted an enhanced performance of perovskite based photovoltaics upon 50\% Sn doping~\cite{RSC_MAPbSnI3-Kanhere}. MAPb$_{0.5}$Sn$_{0.5}$I$_3$ is found to have highest short-circuit photocurrent and broadest light absorption and hence promises highest light harvesting properties among all~\cite{JACS_Feng_Sn-Pbmixed}. Although several researchers have explored the defect physics in pristine MAPbI$_3$~\cite{Yin-APL-defects, Wang_selfdope-APL, JPCL-defect-Du}, but the possibility of Pb-vacancy (Pb-$\Box$) formation, while substituting Sn-atoms, has not been taken into consideration yet. Also, it should be noted that Sn substitution itself can create new defect states in addition to the vacancy states. However, identification and hence controlling the effect of different defects (viz. substitution or Pb-$\Box$) is quite difficult and very much indirect experimentally. An indigenous combination of several experimental tools is mandatory for this, which undoubtedly is rare. This indeed is the reason behind several controversies between the experimentally and theoretically predicted structural parameters of MAPb$_{1-X}$Sn$_X$I$_3$. More importantly, the issues related to defect formation energy, its concentration, and charge state, etc. are still not clearly understood for this rigorously studied semiconducting material. Hall measurements with pristine MAPbI$_3$ film (precursor 1:1 i.e. PbI$_2$:MAI=1:1) suggests that it is n-type self doped material ~\cite{Wang_selfdope-APL}, while Pb-$\Box$s are held responsible for it's p-type behaviour~\cite{Yin-APL-defects}. On the other hand it's experimentally demonstrated that increasing the Sn fraction in the mixed perovskite MAPb$_{1-X}$Sn$_X$I$_3$ can change its oxidation state and thereby changing the material from n-type to p-type~\cite{SciRep_MAPbSnI3_Liu, Wang_selfdope-APL}. To unravel the origin of this change in behaviour of MAPb$_{1-X-Y}$Sn$_X$$\Box_Y$I$_3$ from n-type to p-type, it is therefore mandatory to understand the specific role (i.e. charge state and concentration) of Pb-$\Box$ and the Sn-substitution in controlling the electronic structure. %%%%%%%%%%
\\ \indent 
In this Letter, using state-of-the-art first-principles based methodology under the framework of density functional theory (DFT)~\cite{PRB_Hohen_Kohn-1964, PRB_Hohen_Kohn-1965}, we present a thorough theoretical understanding of the defect physics to explore the role of structural defects viz. Sn substitution and Pb-$\Box$ in regulating the stability and carrier concentration of MAPb$_{1-X-Y}$Sn$_X$$\Box_Y$I$_3$. We find that the different charge states of respective Pb-$\Box$ and substituted Sn play crucial role in determining the electronic structure and hence the carrier concentration. From calculated energetics, as a function of both Sn substitution and Pb-$\Box$ formation, we find that at low temperature Pb-$\Box$ is preferred, while on increasing temperature beyond 250~K Pb-$\Box$ gets destabilized. All our results are duly validated w.r.t experimental evidences from literatures.\\ 
\indent The calculations are performed with all electron based code FHI-aims that uses numeric, atom-centred basis set~\cite{Blum_2009}. A model structure of MAPb$_{1-X-Y}$Sn$_X$$\Box_Y$I$_3$ is created and the supercell size is kept on increasing until the single defect state becomes fully localized inside the supercell~\footnote{Our converged supercell contains 96 atoms (MA$_8$Pb$_8$I$_{24}$). The structures of all different compositions having different number of vacancies and Sn were fully relaxed upto 0.001 eV/${\textrm{\AA}}$ force minimization using the Broyden-Fletcher-Goldfarb-Shanno (BFGS) algorithm. The total energy tolerance is set at 0.0001 eV. The k-mesh convergence is also thoroughly tested and the results that are reported here are performed with 8$\times$8$\times$8 k-mesh size. Tier2 basis set is used with tight settings as implemented in FHI-aims~\cite{Blum_2009}.}. Within the operational temperature range of solar cells (i.e.~$\sim$~320-350~K), MAPbI$_3$ posses a cubic phase~\cite{NatCom-2017-TwinDom}. Also, Sn introduction is supposed to change the crytallographic phase of the perovskite even at room temperature from tetragonal MAPbI$_3$ to cubic MASnI$_3$~\cite{InorgChem_MAPbSnI3_Constantinos}. Thus, we have constrained our calculations to the mostly studied cubic phase in the present study.
In order to ensure that our findings are not just an artifact of the DFT exchange and correlation (XC), as a first step, the DFT XC functionals are thoroughly benchmarked \footnote{The complete comparison of all different functionals i.e., the local-density approximation (LDA)~\cite{PRB_LDA-1992} and several variants of the generalized gradient approximation [PBE, PBEsol~\cite{PRL_PbeSol_2008}, and PW91] against the more advanced hybrid functional HSE06 (that incorporates the non-local exact exchange) will be published elsewhere.}. We find, with our all electron calculations, the experimental band gap ($\sim$ 1.54 eV)~\cite{InorgChem_MAPbSnI3_Constantinos} is very well reproduced by local density approximation (LDA~\cite{PRB_LDA-1992}) [LDA band gap = 1.55 eV]; while both generalized gradient approximation (GGA with PBE)~\cite{GGA_PRL_1997} and hybrid functional [HSE06]~\cite{HSE06} overestimate it by 0.15 and 0.35 eV respectively. The LDA energetics is further validated in Fig.\ref {func}. Here we have demonstrated both the theoretical and experimental~\cite{Nat_PL_Michele} optical spectra  of MAPbI$_3$. The state-of-the-art theoretical spectroscopy techniques as explained in Ref~\cite{PRL_Rinke_MgO} has been employed here (see Fig.~\ref{func} caption for details). The experimental PL peak (corresponding to emission at 1.57 eV) is nicely reproduced by LDA within an error bar of 0.02 eV. \\
\indent Note that this material has its application around 300-350~K temperatures under sunlight. Therefore finite temperature effect is duly included in our theoretical calculations involving lattice thermal vibration at a given temperature. In order to capture the free energy of formation in the working temperature range of solar cells ($\sim$ 300-350~K), the DFT energetics (LDA) is added with the vibrational free energy. The phonon vibration is calculated within the harmonic approximation using the PHONOPY code for the post-processing of the harmonic force constants generated by the finite displacement method. \\
\indent The formation energy of the mixed perovskite MAPb$_{1-X-Y}$Sn$_X$$\Box_Y$I$_3$ (for a 2$\times$2$\times$2 supercell i.e. MA$_8$Pb$_{8-x-y}$Sn$_x$$\Box_y$I$_{24}$) is calculated from the difference of total energies of its precursor materials PbI$_2$, MAI, SnI$_2$ and I$_2$ using the following formula:
\begin{equation}
\begin{split}
E_f(x,y)=\textrm{E}(\textrm{MA}_8\textrm{Pb}_{8-x-y}\textrm{Sn}_x\Box_y\textrm{I}_{24})-8\textrm{E}(\textrm{MAI})\\-x\textrm{E}(\textrm{Sn}\textrm{I}_2)-(8-x-y)\textrm{E}(\textrm{PbI}_2)-y\textrm{E}(\textrm{I}_2)
\label{eq1}
\end{split}
\end{equation}
The co-efficient of each term on the right hand side is so chosen that they stoichiometrically balance the number of MA, Pb, Sn and I atoms in
 MA$_8$Pb$_{8-x-y}$Sn$_x$$\Box_y$I$_{24}$. In order to include the effect of temperature on the formation energy E$_f$ ($x,y,T$), with the total energies of the respective precursors, the free energy of vibration (F$_{vib}$(T)) is added using the formula below:
\begin{equation}
\begin{split}
E_f(x,y,T)=E_f(x,y) + F_{vib}(T) - TS^{config}
\label{eq2}
\end{split}
\end{equation}
The last term of the above equation is known as configurational entropy. It comes from the arrangements of defects in the supercell and is dependent on the total number of defect sites N$_ \textrm{tot}$ and total number of degenerate defect sites N$_\textrm{S}$. 
\begin{equation}
\begin{split}
S^{config}=\textrm{k}_\textrm{B}\frac {\textrm{N}_\textrm{tot}!} {(\textrm{N}_\textrm{S}!(\textrm{N}_\textrm{tot}!-\textrm{N}_\textrm{S}!))}
\label{eq4}
\end{split}
\end{equation}

%\section{Result and Discussion}

\begin{figure}[h!]
\includegraphics[width=0.75\columnwidth,clip]{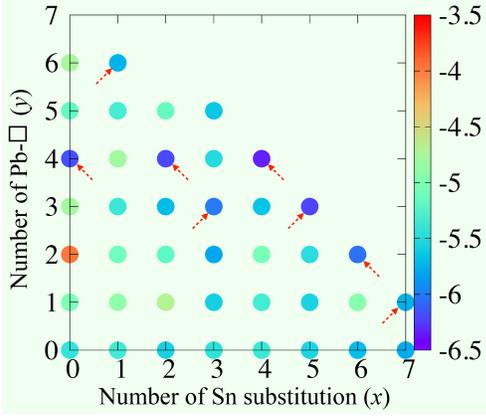}
\caption{Formation energy [$E_f(x,y)$] of different MA$_8$Pb$_{8-x-y}$Sn$_x$$\Box_y$I$_{24}$ structures are represented as a function of Sn substitution and Pb-$\Box$. $E_f(x,y)$ scale is presented by the color-bar in eV. A guide to the eye for the isomers with minimum $E_f(x,y)$ corresponding to each Sn substitution is marked with red arrows.}
\label{Ef-3D}
\end{figure}
\begin{figure}[t!]
\includegraphics[width=1.0\columnwidth,clip]{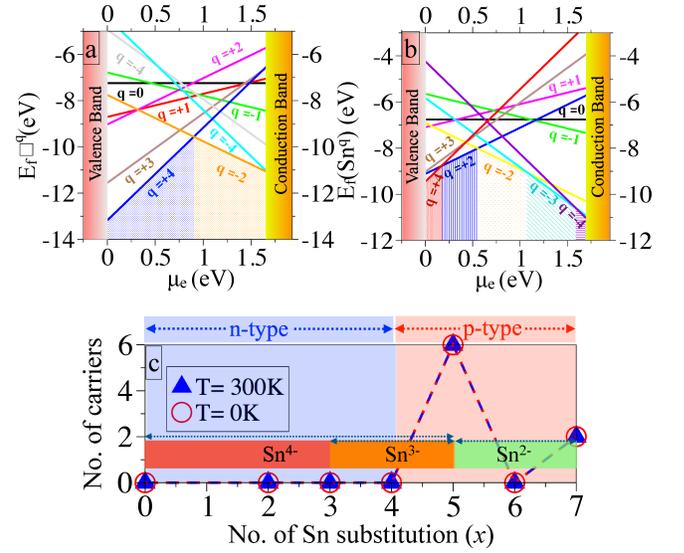}
\caption{The formation energy of single (a) Pb-$\Box$ and (b) Sn substitution as a function of electron chemical potential $\mu_e$ for different charge states q. (c) The carrier concentration of the most stable structures shown with red arrows as in Fig.~\ref{Ef-3D} as a function of Sn substitution and T=0, 300~K.}
\label{chg-plt}
\end{figure}

\begin{figure}[t!]
\includegraphics[width=1.05\columnwidth,clip]{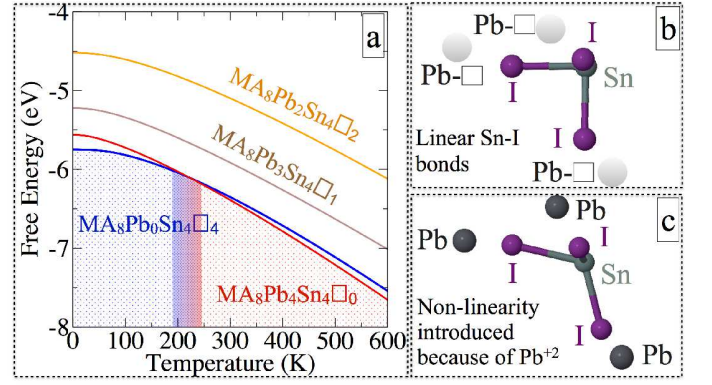}
\caption{(a) Free energy as a function of temperature for different $X$=0.5 isomers; The changes in I-Sn-I bonds in presence of (b) Pb-$\Box$ and (c) Pb atoms.}
\label{phonon}
\end{figure}

 
In Fig.~\ref{Ef-3D} we have plotted the $E_f(x,y)$ for all different possible values of $x,y$. In this regard, we have employed an iterative strategy~\cite{PRB_AB-2016}: at first, we have identified energetically the most stable vacancy (substitution) site in the pristine formula unit MA$_8$Pb$_8$I$_{24}$. Following this, we start scanning over all the other remaining 7 Pb sites to find out the next stable vacancy (substitution) site and so on. 
The process is repeated systematically to increase the number of defects in the system. In  Fig.~\ref{Ef-3D} the most stable isomers with and without the presence of Sn in the lattice are marked with red arrows. The pristine system prefers 4 Pb-$\Box$s per formula unit when there is no Sn substitution. The first introduction of Sn (i.e $x$=1) makes the system energetically unstable and the number of vacancies increases in the system. For $x$=2 and 4, the system prefers 4 vacancies. However, 3 Pb-$\Box$s are preferred when $x$ is odd i.e. $x$= 3 and 5. From the color-code we note that MA$_8$Pb$_0$Sn$_4$$\Box_4$I$_{24}$ is the most stable isomer amongst all different values of $x,y$. It is to mention here that this stable most perovskite structure with $X$=0.5 (i.e. $x$=4) is also the most promising light-harvesting material among the family because of it's lowest band-gap and higher photovoltaic performance parameters viz. short-circuit photocurrent density, open-circuit voltage etc~\cite{JACS_Feng_Sn-Pbmixed}.\\
\indent It's, therefore, needed to understand next, why certain number of Pb-$\Box$ ($y$) are preferred for a given value of Sn substitution ($x$). In neutral MA$_8$Pb$_{8-x-y}$Sn$_x$$\Box_y$I$_{24}$, the role of MA$_8$ is to donate 8 electrons ($e^-$s) in the system. For a specific value of $x$, the system prefers to get stable with certain no. of Pb-$\Box$ ($y$) when this 8 $e^-$s are nicely compensated in the perovskite structure. In view of this, it is important to know the preferred charge state ($q$) of a single defect in the empty inorganic perovskite cage without MA$_8$ (i.e. Pb$_7$Sn$_1^q$I$_{24}$ and Pb$_7$$\Box_1^q$I$_{24}$). To do that the formation energy of single defect at a given charge state ($q$)  is calculated as a function of electron chemical potential ($\mu_e$) [see Fig.~\ref{chg-plt}]. The single defect formation energy at various charge states ($q$) is calculated keeping the pristine neutral Pb$_8$I$_{24}$ as the reference state~\cite{RevModPhys_VanDeWalle-2014, PRB_AB-2016}:\\
Formation energy for Pb-$\Box$:
\begin{equation}
\begin{split}
E_f(\Box^q) = \textrm{E(Pb}_7\Box_1^q\textrm{I}_{24}) - \textrm{E(Pb}_8\textrm{I}_{24}) + \textrm{E(PbI}_2)\\ - \textrm{E(I}_2) + q(\mu_e + \Delta  \textrm{V}^\Box + \textrm{VBM})
\end{split}
\end{equation}
and the same for Sn-substitution:
\begin{equation}
\begin{split}
E_f( \textrm{Sn}^q) =  \textrm{E(Pb}_7 \textrm{Sn}_1^q \textrm{I}_{24}) -  \textrm{E(Pb}_8 \textrm{I}_{24}) +  \textrm{E(PbI}_2)\\ -  \textrm{E(SnI}_2) + q(\mu_e + \Delta \textrm{V}^{\textrm {Sn}}  +  \textrm{VBM})
\end{split}
\end{equation}
$\Delta$V$^{\Box}$ and $\Delta$V$^{\textrm {Sn}}$  represent the core level alignment of E(Pb$_7$$\Box_1^q$I$_{24}$) and E(Pb$_7$Sn$_1^q$I$_{24}$) w.r.t pristine neutral Pb$_8$I$_{24}$; $\mu_e$ represents the electronic chemical potential w.r.t the valance band maxima (VBM) of pristine neutral Pb$_8$I$_{24}$ perovskite.
From Fig.~\ref{chg-plt}, we see that while near the conduction band minima (CBm)  Pb-$\Box$ prefers -2 charge state, it takes +4 when the fermi energy lies near the VBM (see Fig.~\ref{chg-plt}a). However, for Sn substitution, a broader distribution of charge states is observed. -4, -3, -2, +2, +4 charge states are preferred respectively as one moves the fermi level from CBm towards VBM (see Fig.~\ref{chg-plt}b). 
%%%%
%%%%
By comparing the formation energies of a single defect in Fig.~\ref{chg-plt} (a) and (b), we note that near CBm (i.e. n-type doped) Sn substitution is energetically more favoured than Pb-$\Box$ formation. However, if the fermi energy lies near the VBM (i.e. p-type doped), the system behaves in exactly opposite manner; i.e. it prefers Pb-$\Box$ over Sn substitution. We have then computed the net amount of carrier ($e^-$ or hole ($e^+$)) concentration of all the most stable configurations (marked with red arrows in Fig.~\ref{Ef-3D}) by integrating their respective density of states near CBm or VBM respectively. Fig.~\ref{chg-plt}c represents the net carrier concentration of those structures both at 0~K (DFT structure) and at 300~K. We see in Fig.~\ref{chg-plt}c that except for Ma$_8$Pb$_0$Sn$_5\Box_3$I$_{24}$ and Ma$_8$Pb$_0$Sn$_7\Box_1$I$_{24}$, none of the isomers have any additional free carriers in the conduction band or valence band.\\
\indent The above analysis now helps us in understanding the extra stability for a specific value of $x,y$ in MA$_8$Pb$_{8-x-y}$Sn$_x\Box_y$I$_{24}$. 
From Fig. \ref{Ef-3D}, for $x=0$, we see that the most stable phase is $y$=4 i.e. MA$_8$Pb$_4$$\Box_4$I$_{24}$. In MA$_8$Pb$_4$$\Box_4$I$_{24}$, 8$e^-$s, donated by 8 MA$^+$, are perfectly balanced by the 4 Pb-$\Box$s (each Pb-$\Box$ takes 2$e^-$s) leaving no extra carrier in the system. From it's density of states, we note that the system behaves as an n-type semiconductor (as shown in Fig.~\ref{chg-plt}c for $x=0$) and therefore a Pb-$\Box$ should take 2$e^-$s (see Fig.~\ref{chg-plt}a). For $x=2$ we have $y=4$, i.e. in MA$_8$Pb$_2$Sn$_2$$\Box_4$I$_{24}$ when 2 Sn atoms substitute 2 Pb atoms in the lattice, 2 Pb-$\Box$s (2$\times$2$e^-$=4$e^-$s) electronically balance 1 Sn (4$e^{-}$s). Out of the rest 8$e^-$s (coming from 8 MA$^+$), 4 are taken by 1 Sn (4$e^{-}$s) and other 4 are balanced by remaining two Pb-$\Box$s (2$\times2$$e^{-}$=4$e^-$s). Increasing the number of Sn substitution in lattice ($x$=3), reduces the number of Pb-$\Box$ by 1 i.e. $y$=3 forming MA$_8$Pb$_2$Sn$_3$$\Box_3$I$_{24}$. Here 2 Pb-$\Box$ (2$\times$2$e^-$=4$e^-$s) get electronically balanced by 1 Sn (4$e^-$s). Remaining 8$e^-$s donated by 8 MA$^+$ are balanced by 1 Pb-$\Box$ (2$e^-$) and 2 Sn (2$\times$3$e^{-}$=6$e^-$). Likewise, for $x=4$ the most stable configuration is MA$_8$Sn$_4$$\Box_4$I$_{24}$ (i.e. $y=4$, see Fig.~\ref{Ef-3D}). While 4 Pb-$\Box$ (4$\times$2$e^-$=8$e^-$s) and 2  Sn (2$\times$4$e^{-}$=8$e^-$s) are electronically balanced, the residual 2  Sn atoms (2$\times$4$e^-$=8$e^-$s) take care of the 8 electrons coming from 8 MA$^+$s. As one further increases the Sn substitution ($x \textgreater$4), this hybrid perovskite gets modified from n-type to p-type (see Fig.~\ref{chg-plt}c) i.e. the fermi level gets shifted towards VBM. This observation is inline with experimentally observed fact that increasing the Sn concentration beyond 50\% makes the mixed perovskite p-type~\cite{SciRep_MAPbSnI3_Liu}. This change consequently affects the preferred charge states of the respective defects as well. While the system is p-type doped, the preferred charge state of Sn is mixed. It gradually gets modified to Sn$^{2-}$ from Sn$^{4-}$ via Sn$^{3-}$ (see the values near CBm and VBM respectively in Fig.~\ref{chg-plt}b). Note that for $x$=5, we have the most stable case for $y$=3 (i.e. MA$_8$Pb$_0$Sn$_5$$\Box_3$I$_{24}$). Here Sn prefers -2 charge state and this additional Sn substitution takes place at the cost of one Pb-$\Box$ (-2) because they are electronically equal. While 2 of the Pb-$\Box$s (2$\times$2$e^-$=4$e^-$) electronically balance 1 Sn (4$e^-$) in MA$_8$Pb$_0$Sn$_5$$\Box_3$I$_{24}$, 8$e^-$ donated by 8 MA$^+$ are taken by 1 Pb-$\Box$ (2$e^-$) and 2 Sn (2$\times$3$e^{-}$=6$e^-$). A deficit of 6$e^-$ (i.e. presence of 6$e^+$) is generated due to the presence of the remaining 2 Sn atoms that can still accommodate (2$\times$3$e^{-}$=6$e^-$) extra 6$e^-$. This is shown in Fig.~\ref{chg-plt}c for the system MA$_8$Pb$_0$Sn$_5$$\Box_3$I$_{24}$ where the carrier concentration is noted to be 6$e^+$. On increasing the Sn content further to $x$=6 as in MA$_8$Pb$_0$Sn$_6$$\Box_2$I$_{24}$ two of the Sn atoms (2$\times$2$e^{-}$=4$e^-$) completely balances 2  Pb-$\Box$ (2$\times$2$e^{-}$=4$e^-$), while the 8$e^-$ (from the organic moieties) are further completely balanced by 4 Sn atoms (4$\times$2$e^{-}$=8$e^-$). And finally for $x$=7, we note that the value of $y$=1 (i.e. MA$_8$Pb$_0$Sn$_7$$\Box_1$I$_{24}$). This material is very much p-type in nature and thereby the position of fermi-level being close to VBM, Pb-$\Box$ now donates 4$e^-$ along with 8$e^-$ from the organic moieties. The entire 12$e^-$s are completely balanced by 6 Sn atoms (6$\times$2$e^-$=12$e^-$). The remaining single Sn can take two extra $e^-$s and thereby making the carrier concentration to be 2$e^+$.\\
%%%%%%%%%%%%%%%%%%%%%%%%%%%%%%%%%%%%
\indent Further, we have checked the carrier concentrations of these stable structures at T=300~K to ensure the validity of our above analysis at working temperatures. We have taken the structures after 8ps of molecular dynamics simulation at 300~K with Nose-Hoover thermostat. As we can see from Fig.~\ref{chg-plt}c, no changes in the carrier concentration are noted at this elevated temperature. 
It is indeed very crucial to check if the DFT-predicted stable isomers at 0~K remain the most stable ones at higher temperatures or not. Figure~\ref{phonon} depicts the variation of formation energy as a function of temperature following equation~\ref{eq2} for $x$ = 4 isomers. We can see that even after taking into account the vibrational energy, MA$_8$Pb$_0$Sn$_4$$\Box_4$I$_{24}$ still is the energetically most favoured isomer upto $\approx$~195~K. Afterwards, it starts competing with the next isomer MA$_8$Pb$_4$Sn$_4$$\Box_0$I$_{24}$ and they both remain equally probable till $\approx$~250~K. At very higher temperatures, however, the system starts preferring lesser vacancies. We understand from the structural differences between MA$_8$Pb$_0$Sn$_4$$\Box_4$I$_{24}$ and MA$_8$Pb$_4$Sn$_4$$\Box_0$I$_{24}$ (Fig.~\ref{phonon} b, c ) that the presence of Pb hinders the linearity in the system (i.e. Sn-I bonds become non-linear). That is why vacancy is more preferred by the perovskite at $X$ = 0.5 to conserve the linearity at low temperatures. However at higher temperatures, due to thermal disorder, Sn-I bonds become non-linear and therefore presence of Pb is preferred over vacancy beyond $\approx$~250~K. \\
\indent In summary, we have studied the stability characteristics of different stoichiometries in CH$_3$NH$_3$Pb$_{1-X-Y}$Sn$_X$$\Box_Y$I$_3$ perovskites from first principles based calculations under the framework of DFT. The chosen exchange correlation functional is thoroughly benchmarked and validated w.r.t available experimental reports. Our results demonstrate that the system does not prefer Pb to co-exist with Sn, for the most stable configuration at $X$=0.5. On inclusion of the finite temperature effect involving lattice thermal vibration at a given temperature, we find that above 250~K, the Pb-$\Box$s become unfavourable i.e. Pb and Sn occupy equal sites in a formula unit. At higher temperatures, the introduction of thermal disorder, affects the linearity of the Sn-halogen bonds and thereby the system prefers the presence of Pb over Pb-$\Box$s.  We further note that for n-type host the Sn substitution is more preferable than Pb-$\Box$ formation, while for p-type host the Pb-$\Box$  is favoured. Therefore, the gradual increase in Sn content transforms the perovskite from n-type to a p-type semiconductor. This observation is in line with that of Wang et al.~\cite{Wang_selfdope-APL} and Yin et al.~\cite{Yin-APL-defects}, who have shown that Pb-$\Box$s make the system p-type. Finally, we conclude from detailed electronic structural analysis that the preferred charge states of both Pb-$\Box$ and substituted Sn are dependent on the kind of doping of the material. These charge states play crucial role not only in determining the most stable configurations but also to estimate their carrier concentrations at any particular Sn content in the material.\\
\indent We thank IIT Delhi HPC facility for computational resources. We acknowledge the financial support from YSS-SERB research grant, DST, India (grant no. YSS/2015/001209). SB acknowledges Dr. Amrita Bhattacharya for many helpful discussions.
 
% \bibliography{planar}{}

\end{document}